\documentclass[proof]{WileyASNA-v1}

\usepackage{moreverb}

\articletype{Article Type}

\usepackage{epsfig}
\usepackage{multicol}
\usepackage{array}
\usepackage{amsfonts,amsmath,amssymb}
\usepackage{graphicx}

\newcommand\BibTex{{\rmfamily B\kern-.05em \textsc{i\kern-.025em b} \kern-.08em T\kern-.1667em\lower. 7ex\hbox{E}\kern-.125emX}}

\received{......December, 2018}
\revised{.....January, 2019}
\accepted{.........,2019}

\begin{document}

\title{Radiation intensity and polarization from  accretion discs with progressive increasing height}

\author[1]{N. A. Silant'ev*}
\author[1]{G. A. Alekseeva}
\author[1]{V. V. Novikov}

\authormark{ N. A. SILANT'EV  \textsc{et al.}{\normalsize Radiation  intensity and polarization from  accretion discs with progressive increasing height}}

\address[1]{\orgdiv {Central Astronomical Observatory at Pulkovo }, \orgname {}, \orgaddress{\state{St. Petersburg 196140}, \country{Russia}}}

\corres{* N.A.Silant'ev, Pulkovo Observatory.\email{nsilant@bk.ru}}

\presentaddress{Central Astronomical Observatory at Pulkovo , St. Petersburg 196140, Russia}

\abstract[Abstract]{{\bf The paper considers the optically thick accretion discs with the progressive increasing  height. The surface is assumed  to be the conical form.
The radiation with considered wavelength emerges from an ring on the cone and is described by the Milne problem for the intensity and  polarization. 
  The inclination angles  of the rings  are taken 15 and 30 grad. The inclination angles between the line of sight  and  the normal  to the central accretion disc plane are taken  30, 45, 60 grad for every value of the ring inclination. For the continuum radiation the polarization of the emerging light is less than that in the case of the plane accretion disc. The polarization position angles of radiation emerging from the right and left parts of the ring are opposite one another. They are determined by the geometry of the problem.
The position angle of the observed continuum radiation is parallel to the central plane of the accretion disc. 
 Our theory gives the new explanation of that the position angles in "red" and   "blue" wings of a spectral line are opposite one another. This behavior exists in many Seyfert galaxies.}}




 \maketitle

\section{INTRODUCTION}

Frequently one considers  the  plane accretion discs. Sometimes,  the authors  mention that in general case the discs can have the progressive increasing geometric height (see, for example, Shakura \& Sunyaev 1973;  Blandford et al. 2001 and Abramowicz \& Fragile 2013).
  Note that the different aspects of  the accretion discs theory are given in  many papers (see, for example,    
Abramowicz, Calvani and Nobili 1980;  Abramowicz et al. 1988 ; Straub et al. 2011; the review Abramowicz \& Fragile 2013 and Vincent et al. 2015). We use the most popular Chandrasekhar's solution of the Milne problem (see Chandrasekhar 1960). According to this book, the intensity and polarization of radiation, emerging from the optically thick plane atmosphere, are determined by the  radiation diffusion from the sources of thermal radiation located far below the atmospheric surface. 

 The observed intensity $I(\mu)$ and  polarization Stokes parameter $Q(\mu)$ depend on $\mu=\cos\theta={\bf n\cdot N}$, where ${\bf n}$ is the line of sight and ${\bf N}$ is the normal to the surface of an atmosphere. The intensity $I(\mu)$, emerging  along the normal ${\bf N}$,  is $\sim3$ times larger than that going parallel to the surface. The degree of polarization $p(\mu)=Q(\mu)/I(\mu)$ is equal to zero for the direction ${\bf n}\parallel {\bf N}$  ($\mu=1$) and takes the maximum  value $11.73$\% for ${\bf n}\perp {\bf N}$ ($\mu=0$).  The wave electric field oscillations ${\bf E}$ are parallel to the  surface of a disc, namely ${\bf E}\parallel {\bf [ n N ]}$.  Recall, that we consider the Milne problem describing the multiple scattering of continuum radiation in the conservative atmosphere.   In such atmosphere the polarization degree is independent of the wavelength $\lambda$.

The generalized Milne problem presents the angular distribution and polarization degree of the radiation, escaping from the optically thick  scattering atmosphere with  an absorption (see Silant'ev et al. 2017b). In such atmosphere the intensity $I(\mu)$ is more elongated  and the maximum polarization at $\mu=0$ is greater than that in conservative case.

 The angular distribution $J(\mu)=I(\mu)/I(0)$ and the polarization degree $p(\mu)=Q(\mu)/I(\mu)$  are given in Tables 1 and 2 with the interval $\Delta\mu=0.005$. The value $I(0)$ is proportional to the flux of radiation (see Chandrasekhar 1960).  Note that $Q(\mu)=J(\mu)p(\mu)I(0)$. Thus, the angular distribution $J(\mu)$ and $p(\mu)$ are equivalent to the parameters $I(\mu)$ and $Q(\mu)$.

  For the plane accretion disc  the observed polarization $p(\mu)$ (see Tables 1 and 2) allows us to estimate the inclination angle $\theta \equiv i$, i.e. the angle between the line of sight ${\bf n}$ and the normal ${\bf N}$ to the accretion disc (see Fig.1). 
 The observed continuum radiation with the wavelength $\sim\lambda$ emerges from the comparatively narrow ring with the radius $r_{\lambda}$,  corresponding to  the Wien displacement  law: $\lambda(cm)T(K)= 0.29$ (see  Joos 1986).
Below we show that the total radiation polarization from a inclined ring is smaller than in the case of Milne's problem for the plane surface. This polarization is different for the different rings, depending on the inclination angles of them.

If the polarization degrees for different wavelengths are  different, then the idea arises that the corresponding inclinations for the rings are different, i.e. the accretion disc is not plane. As an example, we take the polarimetric observations of NGC 4258  by  Barth et al. 1999. According to their observations, more polarized radiation corresponds to the radiation with small wavelength $\lambda$, i.e. polarization decreases with the increase of $r_{\lambda}$. Just such behavior takes place for the progressive non-linear increasing height of the accretion disc.

 It appears that  the other explanations exist for the  different polarization degrees, corresponding to different wavelengths.  But the non-linear progressive increase of the height in optically thick disc  appears to be most natural and simple.

Our  aim is to calculate the polarization degree $p$ and position angle $\chi$ of radiation escaping from  every part of inclined ring, i.e. to obtain their dependence on azimuthal angle $\varphi$, which characterizes the position of them on the ring. The angles of inclination $\alpha$ for different rings can be different. We assume that  inside the every ring the  inclination ${\alpha}$ is constant. We consider two values $\alpha=15^{\circ}, 30^{\circ}$.  The inclination angles $\theta$  of the line of sight ${\bf n}$ will be taken $30^{\circ}, 45^{\circ}$ and $60^{\circ}$  for every value of  $\alpha$ (see Fig.1).

\begin{figure}
\centering
\fbox{\includegraphics[width=8cm, height=8cm ]{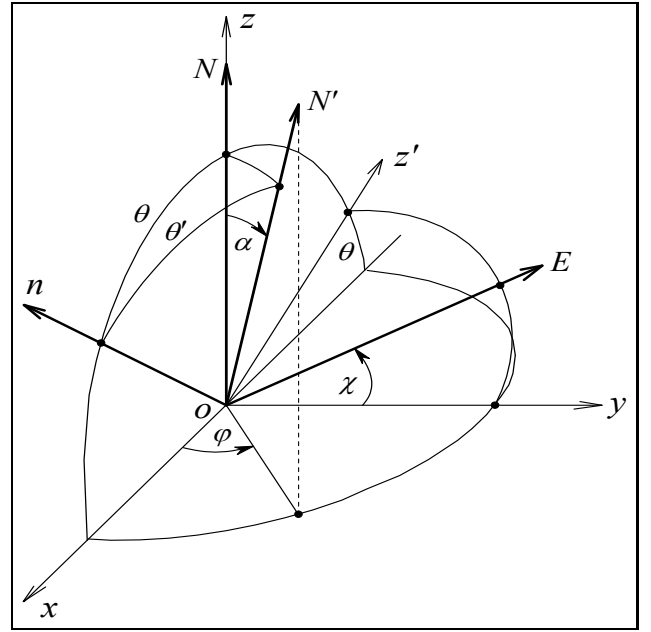}}
\caption{The basic notions}
\label{a}
\end{figure}

\section{ STATEMENT OF PROBLEM}

We know the angular distribution $J(\mu)$ and polarization degree $p(\mu)$ for the conservative Milne problem (see Tables 1 and 2). For the  plane accretion disc the parameter $\mu=\cos\theta={\bf n}{\bf N}$ is identical for all radiating surfaces. Now we consider the conical surface of the ring over the central  plane of an accretion disc. The angle of the cone aperture is $\alpha$, i.e. this is the angle between conical  surface and the central plane of the accretion disc. The normal to the conical surface ${\bf N'}$   is characterized by the angle $\alpha$ and the azimuthal angle $\varphi$ of the point considered on the  surface (see Fig. 1):
\begin{equation}
{\bf N'}={\bf e}_z\cos\alpha-{\bf e}_x\sin\alpha \cos\varphi -{\bf e}_y\sin\alpha\sin\varphi.
\label{1}
\end{equation}
\noindent Here ${\bf e}_z={\bf N},\,\, {\bf e}_x $ and ${\bf e}_y$ are the unit vectors of the common coordinate system $(x,y,z)$. Recall, that ${\bf N}$ is the normal to the central plane of accretion disc. The line of sight (the direction to a telescope) ${\bf n}$ is placed in the plane $({\bf e}_x{\bf N})$:
\begin{equation}
{\bf n}={\bf e}_z\cos\theta+{\bf e}_x\sin\theta.
\label{2}
\end{equation}
\noindent  The angle $\theta$ is the angle between vectors ${\bf n}$ and ${\bf N}$. The direction with the  azimuthal angle $\varphi=0$ coincides with the axis $x$.

The electric wave oscillations ${\bf E}$ of the emerging radiation from optically thick atmosphere are parallel to the vector product  ${\bf [ n N' ]}$ direction:
\[
{\bf A'}\equiv{\bf [ n N' ]}={\bf e}_x\cos\theta \sin\alpha\sin\varphi 
\]
\begin{equation}
 -  {\bf e}_y(\sin\theta\cos\alpha+\cos\theta\sin\alpha\cos\varphi )-{\bf e}_z\sin\theta\sin\alpha\sin\varphi.
\label{3}
\end{equation}
\noindent The direction of ${\bf E}$ - oscillations for the plane accretion disc ($\alpha=0, \varphi=0)$ hold along the direction $({\bf [e_x N]})\sim{\bf e}_y $, i.e. is parallel to the $y$-axis. The electric wave vector ${\bf E}$ is perpendicular to the line of sight ${\bf n}$. It holds in the plane $({\bf e}_{z'}{\bf e}_y)$, where the unit vector ${\bf e}_{z'}$ lies in the plane  $({\bf e}_x{\bf N})$ (see Fig. 1):
\begin{equation}
{\bf e}_{z'}={\bf e}_z\sin\theta+{\bf e}_x\cos\theta.
\label{4}
\end{equation}
\noindent The  reference frame  $({\bf e}_{z'},{\bf e}_y)$ corresponds to the telescope coordinate system.   If the radiating $\varphi$-point lies on the conical surface, then  the ${\bf E}$-vector oscillates along the direction, characterized  by the angle $\chi$  (see Fig.1). It is easy obtain the following formula : 
\begin{equation}
\tan{\chi}=\frac{A'_{z'}}{A'_y}=\frac {\sin{\alpha}\sin{\varphi}}{\sin{\theta}\cos{\alpha}+\cos{\theta}\sin{\alpha}\cos{\varphi}}.
\label{5}
\end{equation}
\noindent For $\varphi=0^{\circ}$ or $\varphi= 180^{\circ}$ the angle $\chi=0$, i.e. the electric oscillations hold along the $y$-axis. For $0^{\circ}\le\varphi\le180^{\circ}$ the angle $\chi\ge0$ and for $180^{\circ}\le\varphi\le360^{\circ}$ the angle $\chi\le0$.

Taking $\varphi =90^{\circ}$, we obtain from Eq.(5) the formula for $\chi_{max}$
\begin{equation}
\tan{\chi_{max}}=\frac{\sin{\alpha}}{\cos{\alpha}\sin{\theta}}.
\label{6}
\end{equation}

 Note that the angle $\chi$  does not depend on the polarization degree $p(\varphi)$ of the radiation, emerging from this point.  The angle $\chi$ depends  on the angles $\alpha,\, \theta $ and $\varphi$, i. e. depends on pure geometrical values. This  is the consequence that the radiating light has the ${\bf E}$ - oscillations parallel to the local radiating surface.   Later we will discuss this effect for the explanation of the position angles  in the  spectral  lines emission from the rotating accretion disc.

 According to Eq.(5), the position angle $\chi$ increases from $\chi=0$ at $\varphi=0$ to the maximum value at $\varphi=90^{\circ}$, and then symmetrically decreases up to zero value at $\varphi=180^{\circ}$. The angle $\chi$  increases with the increasing of the angle $\alpha$.
 Due to the right - left symmetry of the problem, the  observation of all disc in continuum  gives rise to the value $\chi_{total}=0$.

 The Milne problem possesses the axial symmetry. In this case the position angle of the emerging radiation can have two possible directions of the $\bf E $-oscillations - perpendicular to the plane $({\bf nN})$ and in this plane. The second case holds, if the light fluxes parallel to the surface are greater, than that along the normal ${\bf N}$ (see Dolginov et al. 1995).  Usually, the first case is valid. So, formula (5) holds both in the Milne problem and in every another problem with axial symmetry.

Thus, we have the formulas to calculate the position angle $\chi(\varphi,\alpha, \theta)$
  of radiation escaping from the inclined surface.  This angle is determined by the geometrical values - $\alpha, \varphi$ and $\theta$. 

\subsection{Calculation of polarization degree}

According to the Milne problem, the intensity and polarization of the radiation, emerging from  $\varphi$-part of conical ring, depend on the angle between the line of sight ${\bf n}$ and the normal ${\bf N'}$ (see Fig.1), i.e they depend on $\mu'={\bf n\cdot N'}$.  To obtain these  values, we must calculate the parameter $\mu' $ and take the values of $J(\mu')$ and $p(\mu')$ from Tables 1 and 2. The parameter $\mu'$ can be calculated from the following formula
\begin{equation}
\mu'(\alpha,\theta,\varphi)={\bf n\cdot N'}=\cos\theta\cos\alpha-\sin\theta\sin\alpha\cos\varphi.
\label{7}
\end{equation}
\noindent Note that $ \mu'(\alpha,\theta,\varphi)\equiv\mu'(\alpha,\theta,-\varphi)$, i.e. the intensity and polarization degree are the  symmetric functions of the azimuthal angle $\varphi$.

 A telescope observes the radiation fluxes $F_I$  and $F_Q$ from an accretion disc. For the plane accretion disc the flux $F_I({\bf n})$ in the direction ${\bf n}$ is
\begin{equation}
F_I({\bf n})=\frac{1}{R^2}\mu \,J(\mu))I(0)S_{ring}.
\label{8}
\end{equation}
\noindent $R$ is the distance to a telescope, $S_{ring}=2\pi \rho\Delta \rho$ is  the total plane ring surface, which emerges the  radiation with wavelength $\lambda$ (remember that  $\lambda(cm)T(K)= 0.29$ ). The factor $\mu={\bf n\cdot N}$ describes the projection of the plane  ring surface perpendicular to the line of sight ${\bf n}$.

The Stokes parameter $F_Q({\bf n})$ is described by the formula
\begin{equation}
F_Q({\bf n})=\frac{1}{R^2}\mu\, J(\mu)I(0)p(\mu)S_{ring}.
\label{9}
\end{equation}
\noindent The degree of the observed polarization $p_F(\mu)=F_Q({\bf n})/F_I({\bf n})=p(\mu)$. The values $J(\mu)$ and $p(\mu)$ are given in Tables 1 and 2. Recall, that  $I(0)$ is proportional to the flux of emerging radiation ( see Chandrasekhar 1960).

 For the inclined ring the parameter $\mu'={\bf n\cdot N'}$ is different for every azimuthal angle $\varphi$. The corresponding small surface element is equal to $d S(\varphi)=\rho\Delta\rho d\varphi$.  The value $\Delta\rho$ is the width of the ring. The flux $F_I({\bf n},\alpha)$ from the inclined  conical ring is equal to
\begin{equation}
F_I({\bf n},\alpha)= \frac{1}{R^2}\int _0^{2\pi} dS(\varphi)\, \mu'J(\mu')I(0).
\label{10}
\end{equation}
\noindent Note that $\mu' $ depend on angles $\varphi, \alpha, \theta $ (see Eq.(7)\,) . For  the value $F_Q({\bf n},\alpha)$  we have
\[
F_Q({\bf n},\alpha)=
\]
\begin{equation}
\frac{1}{R^2}\int_0^{2\pi} dS(\varphi)\, \mu' J(\mu',\alpha,\varphi)\,p(\mu',\alpha, \varphi)
\cos{2\chi}I(0).
\label{11}
\end{equation}
\noindent The integral value $F_U\equiv 0$, due to asymmetric dependence  of the angle $\chi(\varphi)$ on the azimuthal angle $\varphi$: $\chi(-\varphi)=-\chi(\varphi)$ (see Eq.(5) ). Note that the polarization degree  $p(\mu')$ and position angle $\chi(\varphi)$ do not depend on the constant value  $I(0)$.

 Here we used the  transformation formula for the Stokes parameters at the rotation of the axes (see, for example, Chandrasekhar 1960 and  Dolginov et al. 1995). This formula is used  to obtain the parameter $F_Q$ in the telescope reference frame  ($yz'$). The function $\cos{2\chi}$  in Eq.(11)  describes this transformation. The term $\mu'$ describes the projection of the  inclined small surface element $dS(\varphi)$  perpendicular to the  line of sight ${\bf n}$.

\begin{table}[h!]
\caption {  The angular distribution $J(\mu)$ and degree of polarization $p(\mu)=Q(\mu)/I(\mu)$ in \% for the Milne problem in conservative atmosphere}
\scriptsize
\begin{tabular}{ l | c c | l c c|}
\hline
\noalign{\smallskip}
$\mu$ & $J{(\mu)}$ & $p{(\mu)}$ & $\mu$ & $J(\mu)$ & $p(\mu)$ \\
\hline
\noalign{\smallskip}
 0        & 1           & 11.7128  & 0.255 & 1.5974 & 4.6001 \\
 0.005 & 1.0200 & 11.2307 & 0.260 & 1.6078 & 4.5339 \\
0.010 & 1.0366 & 10.8776 & 0.265 & 1.6181& 4.4687 \\
 0.015 & 1.0519 & 10.5711 & 0.270 & 1.6284 & 4.4046 \\
 0.020 & 1.0664 & 10.2949 & 0.275 & 1.6387 & 4.3415 \\
 0.025 & 1.0805 & 10.0410 & 0.280 & 1.6489 & 4.2795 \\
 0.030 & 1.0941 & 9.8048 & 0.285 & 1.6592 & 4.2183 \\
 0.035 & 1.1074 & 9.5831 & 0.290 & 1.6694& 4.1582  \\
 0.040 & 1.1204 & 9.3737 & 0.295 & 1.6797 & 4.0989 \\
 0.045 & 1.1331 & 9.1750 & 0.300 & 1.6899& 4.0406 \\
 0.050 & 1.1458 & 8.9857 & 0.305 & 1.7001 & 3.9831 \\
 0.055 & 1.1582 & 8.8047 & 0.310& 1.7103 & 3.9265  \\
 0.060 & 1.1704 & 8.6313 & 0.315 & 1.7204 & 3.8707  \\
 0.065 & 1.1822 & 8.4646 & 0.320 & 1.7306 & 3.8158  \\
 0.070 & 1.1945 & 8.3043 & 0.325 & 1.7407 & 3.7616  \\
 0.075 & 1.2064 & 8.1496 & 0.330 & 1.7509 & 3.7082  \\
 0.080 & 1.2182 & 8.0003 & 0.335 & 1.7610& 3.6555  \\
 0.085 & 1.2298 & 7.8558 & 0.340 & 1.7711 & 3.6036  \\
 0.090 & 1.2414 & 7.7160 & 0.345 & 1.7812 & 3.5524  \\
 0.095 & 1.2529 & 7.5805 & 0.350 & 1.7913 & 3.5020  \\
 0.100 & 1.2644 & 7.4489 & 0.355 & 1.8014 & 3.4522 \\
 0.105 & 1.2757 & 7.3212 & 0.360 & 1.8115 & 3.4031  \\
 0.110 & 1.2870 & 7.1970 & 0.365 & 1.8216& 3.3546  \\
 0.115 & 1.2983 & 7.0762 & 0.370 & 1.8316 & 3.3068  \\
 0.120 & 1.3094 & 6.9586 & 0.375 & 1.8417 & 3.2596   \\
 0.125 & 1.3206 & 6.8441 & 0.380 & 1.8517 & 3.2131  \\
 0.130 & 1.3316 & 6.7324 & 0.385 & 1.8618 & 3.3.167  \\
 0.135 & 1.3427 & 6.6235 & 0.390 & 1.8718 &  3.1217  \\
 0.140 & 1.3536 & 6.5172 & 0.395 & 1.8818 & 3.0770   \\
 0.145 & 1.3646 & 6.4134 & 0.400 & 1.8918 & 3.0328  \\
 0.150 & 1.3755 & 6.3120 & 0.405 & 1.9018 &  2.9891  \\
 0.155 & 1.3863 & 6.2128 & 0.410 & 1.9118 & 2.9460  \\
 0.160 & 1.3972 & 6.1159 & 0.415 & 1.9218 & 2.9034  \\
 0.165 & 1.4079 & 6.0211 & 0.420 & 1.9318 & 2.8614  \\
 0.170 & 1.4187 & 5.9283 & 0.425 & 1.9418 & 2.8198  \\
 0.175 & 1.4294 & 5.8375 & 0.430 & 1.9517 & 2.7788  \\
 0.180 & 1.4401 & 5.7486 & 0.435 & 1.9617 & 2.7383  \\
 0.185 & 1.4507 & 5.6614 & 0.440 & 1.9716 & 2.6982  \\
 0.190 & 1.4614 & 5.5760 & 0.445 & 1.9816 & 2.6586  \\
 0.195 & 1.4720 & 5.4923 & 0.450 & 1.9915 & 2.6195  \\
 0.200 & 1.4823 & 5.4102 & 0.455 & 2.0015 & 2.5809  \\
 0.205 & 1.4931 & 5.3296 & 0.460 & 2.0114 & 2.5426  \\
 0.210 & 1.5036 & 5.2506 & 0.465 & 2.0213 & 2.5049  \\
 0.215 & 1.5142 & 5.1730 & 0.470 & 2.0312 & 2.4676  \\
 0.220 & 1.5246 & 5.0969 & 0.475 & 2.0411 & 2.4306  \\
 0.225 & 1.5351 & 5.0221 & 0.480 & 2.0510 & 2.3942  \\
 0.230 & 1.5455 & 4.9487 & 0.485 & 2.0609 & 2.0609  \\
 0.235 & 1.5559 & 4.8765 & 0.490 & 2.0708 & 2.3224  \\
 0.240 & 1.5664 & 4.8057 & 0.495 & 2.0807 & 2.2871  \\
 0.245 & 1.5767 & 4.7360 & 0.500 & 2.0906 & 2.2522  \\
 0.250 & 1.5871 & 4.6675 & 0.505 & 2.1005 & 2.2177  \\
\hline
\end{tabular}

\end{table}
\begin{table}[h!]
\caption {  The angular distribution $J(\mu)$ and degree of polarization $p(\mu)=Q(\mu)/I(\mu)$ in \% for the Milne problem in conservative atmosphere (continue)}
\scriptsize
\begin{tabular}{ l | c c | l c c}
\hline
\noalign{\smallskip}
$\mu$ & $J{(\mu)}$ & $p{(\mu)}$ & $\mu$ & $J(\mu)$ & $p(\mu)$ \\
\hline
\noalign{\smallskip}
 0.510 & 2.1104 & 2.1836 & 0.760 & 2.5993 & 0.8458 \\
0.515  & 2.1202 & 2.1498 & 0.765 & 2.6090 & 0.8248 \\
 0.520 & 2.1301 & 2.1164 & 0.770 & 2.6187 & 0.8039 \\
 0.525 & 2.1399 & 2.0834 & 0.775 & 2.6284 & 0.7832 \\
 0.530 & 2.1498 & 2.0507 & 0.780 & 2.6381 & 0.7627 \\
 0.535 & 2.1597 & 2.0183 & 0.785 & 2.6478 & 0.7423 \\
 0.540 & 2.1695 & 1.9863 & 0.790 & 2.6575 & 0.7221  \\
 0.545 & 2.1793 & 1.9546 & 0.795 & 2.6672 & 0.7021 \\
 0.550 & 2.1892 & 1.9233 & 0.800 & 2.6769 & 0.6822 \\
 0.555 & 2.1990 & 1.8923 & 0.805 & 2.6865 & 0.6625 \\
 0.560 & 2.2088 & 1.8616 & 0.810 & 2.6962 & 0.6430  \\
 0.565 & 2.2187 & 1.8312 & 0.815 & 2.7059 & 0.6236  \\
 0.570 & 2.2285 & 1.8011 & 0.820 & 2.7156 & 0.6043  \\
 0.575 & 2.2383 & 1.7713 & 0.825 & 2.7253 & 0.5852  \\
 0.580 & 2.2481 & 1.7418 & 0.830 & 2.7350 & 0.5663  \\
 0.585 & 2.2579 & 1.7126 & 0.835 & 2.7446 & 0.5475  \\
 0.590 & 2.2677 & 1.6837 & 0.840 & 2.7543 & 0.5288  \\
 0.595 & 2.2775 & 1.6551 & 0.845 & 2.7640 & 0.5103  \\
 0.600 & 2.2873 & 1.6268 & 0.850 & 2.7737 & 0.4917  \\
 0.605 & 2.2971 & 1.5987 & 0.855 & 2.7833 & 0.4737 \\
 0.610 & 2.3069 & 1.5709 & 0.860 & 2.7930 & 0.4557  \\
 0.615 & 2.3167 & 1.5434 & 0.865 & 2.8027 & 0.4377  \\
 0.620 & 2.3265 & 1.5161 & 0.870 & 2.8123 & 0.4199  \\
 0.625 & 2.3363 & 1.4891 & 0.875 & 2.8220 & 0.4022   \\
 0.630 & 2.3461 & 1.4624 & 0.880 & 2.8317 & 0.3847  \\
 0.635 & 2.3558 & 1.4359 & 0.885 & 2.8413 & 0.3673  \\
 0.640 & 2.3656 & 1.4096 & 0.890 & 2.8510 & 0.3500  \\
 0.645 & 2.3754 & 1.3836 & 0.895 & 2.8606 & 0.3328   \\
 0.650 & 2.3851 & 1.3579 & 0.900 & 2.8703 & 0.3158  \\
 0.655 & 2.3949 & 1.3323 & 0.905 & 2.8799 & 0.2989  \\
 0.660 & 2.4046 & 1.3071 & 0.910 & 2.8896 & 0.2822  \\
 0.665 & 2.4144 & 1.2820 & 0.915 & 2.8992 & 0.2655  \\
 0.670 & 2.4242 & 1.2572 & 0.920 & 2.9089 & 0.2490  \\
 0.675 & 2.4339 & 1.2356 & 0.925 & 2.9185 & 0.2326  \\
 0.680 & 2.4437 & 1.2082 & 0.930 & 2.9182 & 0.2163  \\
 0.685 & 2.4534 & 1.1840 & 0.935 & 2.9378 & 0.2001  \\
 0.690 & 2.4631 & 1.1601 & 0.940 & 2.9475 & 0.1840  \\
 0.695 & 2.4729 & 1.1363 & 0.945 & 2.9571 & 0.1681  \\
 0.700 & 2.4826 & 1.1128 & 0.950 & 2.9668 & 0.1523  \\
 0.705 & 2.4924 & 1.0894 & 0.955 & 2.9764 & 0.1366  \\
 0.710 & 2.5021 & 1.0664 & 0.960 & 2.9860 & 0.1210  \\
 0.715 & 2.5118 & 1.0435 & 0.965 & 2.9957 & 0.1055  \\
 0.720 & 2.5215 & 1.0207 & 0.970 & 3.0053 & 0.0901  \\
 0.725 & 2.5313 & 0.9982 & 0.975 & 3.0150 & 0.0748  \\
 0.730 & 2.5410 & 0.9759 & 0.980 & 3.0246 & 0.0596  \\
 0.735 & 2.5507 & 0.9538 & 0.985 & 3.0342 & 0.4460 \\
 0.740 & 2.5604 & 0.9318 & 0.990 & 3.0438 & 0.0296  \\
 0.745 & 2.5701 & 0.9100 & 0.995 & 3.0535 & 0.0148  \\
 0.750 & 2.5798 & 0.8885 & 1        & 3.0631 & 0  \\
 0.755 & 2.5896 & 0.8671 &     -     &    -        &    -           \\
\hline
\end{tabular}
\end{table}

\begin{figure*}
\fbox{\includegraphics[width=16cm, height=8cm]{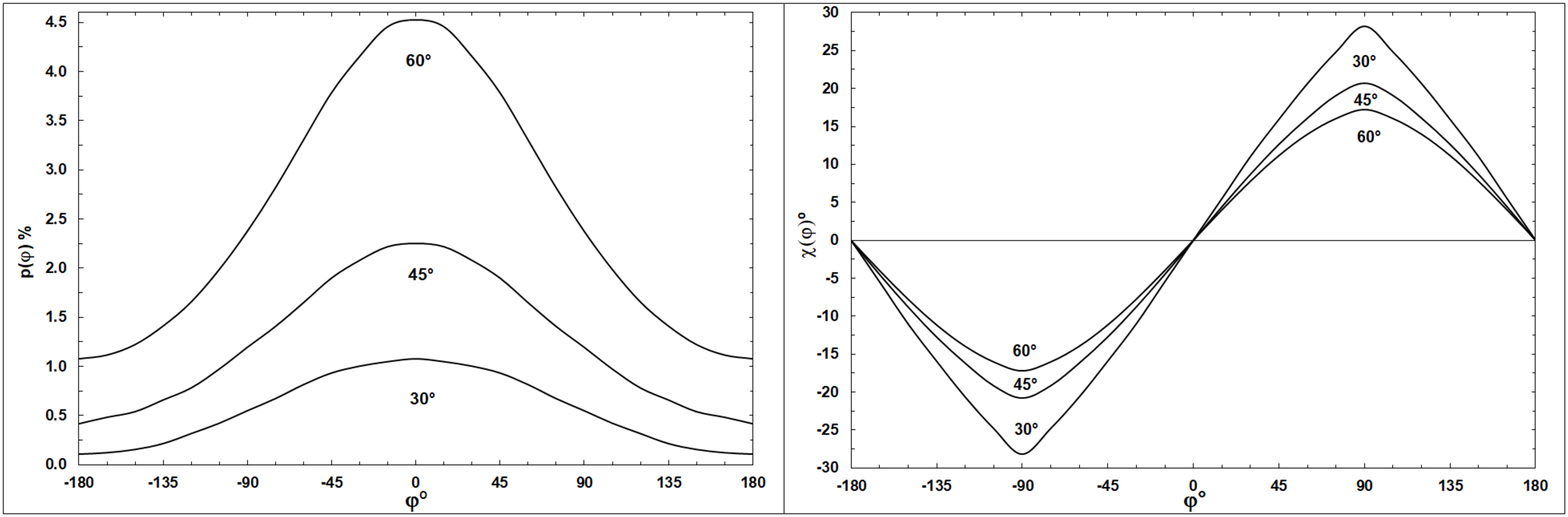}} \caption{Dependence of polarization degree
$p(\varphi)$ in \% and the position angle $\chi(\varphi)$ of the radiation emerging from the $\varphi$ -  position
on inclined ring. The angle of inclination is $\alpha =15^{\circ}$. The numbers denote the inclination angles
$\theta$  of the line of sight.} \label{b}
\end{figure*}

\begin{figure*}
\fbox{\includegraphics[width=16cm, height=8cm]{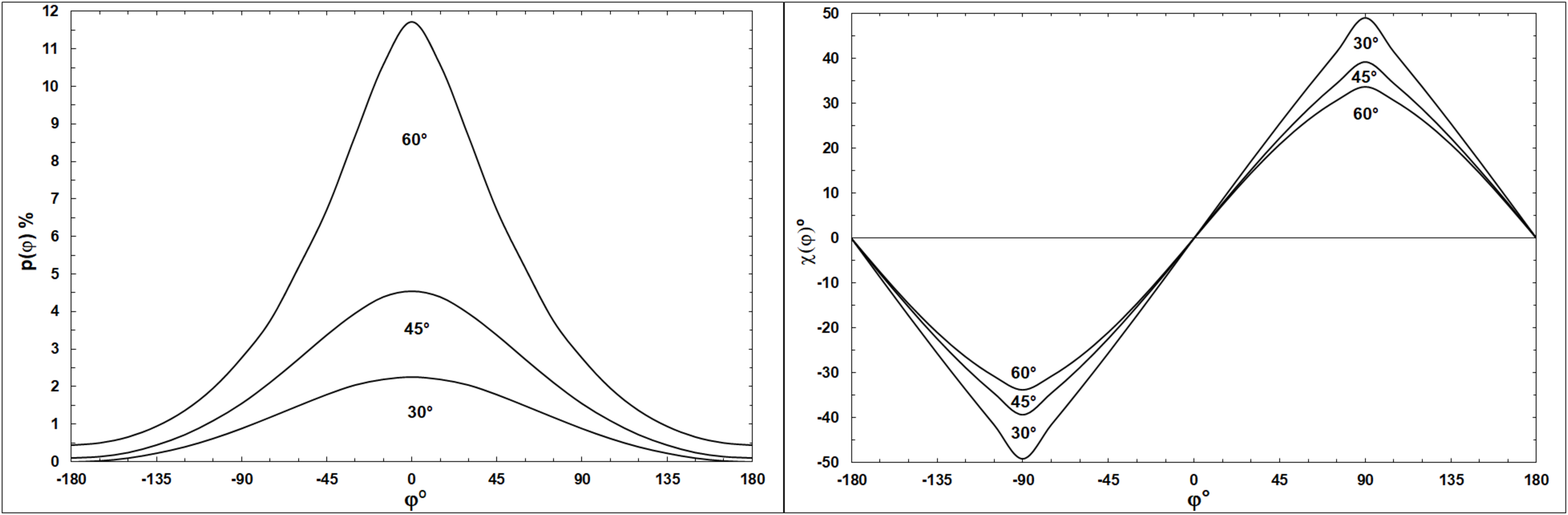}} \caption{Dependence of polarization degree
$p(\varphi)$ in \% and the position angle $\chi(\varphi)$ of the radiation emerging from the $\varphi$ -  position
on inclined ring. The angle of inclination  is $\alpha =30^{\circ}$. The numbers denote  the inclination angles
$\theta$ of the line of sight.} \label{c}
\end{figure*}

\section{RESULTS OF CALCULATIONS}

Our calculations are made  for two ring inclination angles $\alpha=15^{\circ}$ and $30^{\circ}$.  For every $\alpha$ - value we  studied three directions of the line of sight, corresponding to  $\theta=30^{\circ}, 45^{\circ}$ and $60^{\circ}$. Tables 1 and 2  present the angular distribution $J(\mu)$ and the polarization degree  $p(\mu)$  in \% for the Milne problem in optically thick plane electron conservative atmosphere. These Tables have small interval $\Delta\mu =0.005$. This allows us to obtain the $J(\mu')$ and $p(\mu')$ values with  a good accuracy.

In Figs. 2 and 3  we give  the polarization degree $p(\mu')$  and position angle $\chi(\varphi)$ of the  emerging  radiation from the part of conical ring, characterizing by the asimuthal angle $\varphi$. Note that the angles $\chi =0$  correspond to the ${\bf E}$ - oscillations along the $y$-axis (see Fig.1). The intensity is equal to $I(\mu')=J(\mu')I(0)$. The value $J(\mu')$ can be taken from Tables 1 and 2.

In Table 3 we give the values $p_F({\bf n},\alpha)=F_Q({\bf n},\alpha)/F_I({\bf n},\alpha)$ for number of  angles $\alpha$ and $\theta$. They demonstrate that the polarization degree $p_F({\bf n},\alpha)$ diminishes with the increasing of the inclination angle $\alpha$. The polarization increases also with the increasing of the angle $\theta$.

The maximum value of the angle $\chi$ corresponds to the azimuthal angles $\varphi =\pm 90^{\circ}$.   As it was mentioned, the $\chi(-\varphi)=-\chi(\varphi)$, i.e. the  positional angle $\chi$  for the right part of the ring is opposite to that for the left part.  This gives that the observed continuum radiation from all  ring has $\chi_{total}=0$ .
The values $\chi(\varphi)$ diminish with the decrease of the angle $\alpha$.  For $\alpha \to 0$ the value $\chi \to 0$.
 Figs. 2 and 3 demonstrate that the polarization $p(\mu')$   increases with the enhancement  of the angle $\theta$. Such behavior is analogous to the case of the plane accretion disc.

\subsection{Some remarks on spectral line, emerging from the rotating  accretion disc with progressive increasing  height}

In all calculations we used the solution of the Milne problem for electron atmosphere. Strictly speaking, these calculations are not applied for description  of the spectral line radiation transfer. For spectral line the values  $J(\mu,\nu)$ and $p(\mu,\nu)$  differ from those for continuum radiation, given in Tables 1 and 2.  Recall, that the spectral line has  the similar scattering matriсes (for dipole  transition between the radiating levels )(see, for example, Silant'ev et al. 2017a).  Besides, we know that the polarization of the emerging radiation is mostly described by the last scattering before the escaping from the atmosphere. This process is similar for both problems. For these reasons, the qualitative behavior of the spectral line polarization  can be similar to the continuum radiation.

Recall, that the position angle $\chi$ depends on the pure geometrical values - the angles $\varphi, \alpha$ and $\theta$ (see Fig.1). This means that formula (5) is valid for the resonance line emission too.

The observations (see Smith et al. 2002) show that the position angles $\chi$ have the opposite signs in  the "red" and "blue" wings of the $H_{\alpha}$ spectral line.

 The first explanation of this effect is given in Smith et al. 2005. The authors found that the light,  single scattered  in an equatorial torus,  takes opposite position angles for radiation from the right and left parts of rotating disc. This parts, due to the Doppler effect,  give the "red" and "blue" wings.  As a result, the scattered radiation acquires opposite position angles in  the observed "red" and "blue" wings.

Another explanation (see Silant'ev et al. 2013) is related with the azimuthal magnetic field in the rotating accretion disc. The Faraday rotation of the polarized light in the right and left parts gives rise to the opposite position angles.

The new explanation, presented in this paper,  is related with the feature  that the inclined optically thick ring gives opposite position angles for the emerging radiation in the right and left parts of the ring. According to the Doppler effect in rotating accretion disc, these parts corresponds to the "red" and "blue" wings of the spectral line, i.e. they acquire the opposite position angles. This explanation does not use the single scattering of radiation and the existence of azimuthal magnetic field. The effect arises due to the  inclination of a ring,  where the emission of the spectral line radiation holds.

\begin{table}[h!]
\caption {The degree of polarization $p_F$ in \%
of radiation escaping from
a ring with the inclination angles $\alpha$ for the inclination of the line of sight $\theta$.}
\begin{tabular}{l|ccc|}
\hline
\noalign{\smallskip}
&\multicolumn{1}{l}{$\theta=30^{\circ}$}  &
 \multicolumn{1}{l}{$\theta=45^{\circ}$} &
 \multicolumn{1}{l}{$\theta=60^{\circ}$}  \\
\cline{2-2} \cline{3-3} \cline{4-4}
\noalign{\smallskip}
$\alpha^{\circ}$  & $p_F$   & $p_F$ & $p_F$   \\
\noalign{\smallskip}
\hline
\noalign{\smallskip}
0   &  1.06  &  1.89  &  2.35    \\
15 &  0.41  &  0.98  &  1.87   \\
30 &  0.37  &  0.68  &  0.90    \\
  \hline
\end{tabular}
\end{table}

\section{CONCLUSION}

Usually the accretion disc is considered as the plane one. We investigated the case of the optically thick accretion discs with the progressive increasing height,  which is  assumed to be the conical form. We take the ring inclination angle  $\alpha$ aquires the values $15^{\circ}$ and $30^{\circ}$.  For every $\alpha$ we consider three inclination angles for the emerging radiation: $\theta\equiv i=30^{\circ}, 45^{\circ}$ and $60^{\circ}$. The radiation with the considered wavelength $\lambda$ emerges from a ring on the cone and is described by the Milne problem for the intensity and linear polarization. The place, where the $\lambda$-radiation emerges from the surface of a ring, is determined by the Wien displacement  law: $\lambda(cm)T(K)= 0.29$. 
 
  The observed radiation  emerges  from the differently oriented parts of the inclined ring. It was found that the polarization degree  of the observed continuum  radiation flux $p_F$ from the conical surface is  less  than that for the corresponding Chandraselhar value. 

 We demonstrated that the position angles $\chi$ of the emerging radiation from the right and left parts of the inclined ring have opposite signs. This is the consequence that $\chi$ depends on the pure geometrical values - the inclination angle $\alpha$ of the ring, the inclination angle $\theta$ between the line of sight and the normal to the central plane of accretion disc, and on the azimuthal angle $\varphi$ characterizing the place on the ring. The angles $\chi$ diminish with the decreasing of the  inclination angle $\alpha$.

In  a rotating accretion disc the right and left parts of the disc radiate in  the "red" and "blue" wings of the spectral line.
Thus, we give the new mechanism of why the position angles in  the "red" and "blue" wings of the spectral line have the opposite values.
 This behavior is observed in a number of Seyfert active galactic nuclei.
 
{\bf Acknowledgements.}
This research was supported by the Program of Prezidium of Russian Academy of Sciences N 12 " Problems of origin and evolution of the Universe" .
 The authors are very grateful to Prof. Abramowicz for very useful remarks and advices.


\begin{thebibliography}{30}
\bibitem[\protect\citeauthoryear{Abramowicz et al.}{1980}]{b1} Abramowicz, M. A., Calvani, M..Nobili, L., 1980, Astroph. J. 242, 772
\bibitem[\protect\citeauthoryear{Abramowicz et al.}{1988}]{b2} Abramowicz, M. A., Czerny, B., Lasota, J. P., Szuszkiewicz, 1988, Astroph. J. 332, 648
\bibitem[\protect\citeauthoryear{Abramowicz \& Fragile}{2013}]{b3} Abramowicz, M. A., Fragile, P. Ch., 2013, Living Rev. Relativity 16:1, Springer International Publishing (https/:doi.org/10.12942/ lrr-2013-1)
\bibitem[\protect\citeauthoryear{Barth et al.}{1999}]{b4} Barth, A. J., Tran, H. D., Braserton, M. S.. et al., 1999, AJ 118, 1609
\bibitem[\protect\citeauthoryear{ Blandford et al.} {2001}]{b5}  Blandford, R., Broderick, A., Heyl, J., Koopmans, L., Lee, H-W., 2001, in Proceedings of the XII Canary Islands Winter School of Astrophysics, published in ArXiv: astro-ph/0107228
\bibitem[\protect\citeauthoryear{Chandrasekhar}{1960}]{b6} Chandrasekhar, S., 1960, Radiative transfer,  Dover, New York
\bibitem[\protect\citeauthoryear{Dolginov et al.}{1995}]{b7} Dolginov, A. Z., Gnedin, Yu. N., Silant'ev, N. A., 1995, Propagation and Polarization of  Radiation in Cosmic Media. Gordon \&  Breach Publ., Amsterdam, ISBN 2-88124-987-6
\bibitem[\protect\citeauthoryear{Joos}{1986}]{b8} Joos, G., 1986 ,Theoretical Physics, Dover, New York, ISBN 0-486-65227- 0
\bibitem[\protect\citeauthoryear{Shakura \& Sunyaev }{1973}]{b9} Shakura, N. I., Sunyaev, R. A., 1973, A\&A  24, 337
\bibitem[\protect\citeauthoryear{Silant'ev et al.}{2017a}]{b10} Silant'ev, N. A., Alekseeva, G. A., Novikov, V. V., 2017a,  Astrophys. Space Sci., 357, 53
\bibitem[\protect\citeauthoryear{Silant'ev et al.}{2017b}]{b11} Silant'ev, N. A., Alekseeva, G. A., Novikov, V. V., 2017b,  Astrophys. Space Sci., 362, 151
\bibitem[\protect\citeauthoryear{Silant'ev et al.}{2013}]{b12} Silant'ev, N. A., Gnedin, Yu. N., Buliga, S. D., Piotrovich, M. Yu., Natsvlishvili, T. M., 2013,  Astroph. Bulletin  68, 14
\bibitem[\protect\citeauthoryear{Smith et al.}{2002}]{b13}  Smith, J .E., Young, S., Robinson, A, et al. 2004, MNRAS, 335, 773
\bibitem[\protect\citeauthoryear{Smith et al. }{2005}]{b14}  Smith, J. E., Robinson, A., Young, S., Axon, D. J., Corbett, E. A.: 2005,  MNRAS 359, 846
\bibitem[\protect\citeauthoryear{Straub et al.}{2011}]{b15} Straub, O., Bursa, M., Sadowski, A., Steiner, J. F., Abramowicz, M. A., Kluzniak, W., McClintock, J. E., Narayan, R., \& Remillard, R. A., 2011, A\&A, 533, A67 
\bibitem[\protect\citeauthoryear{Vincent et al.}{2015}]{b16} Vincent, F. H., Yan, W., Straub, O., Zdziarski, A. A., Abramowicz, M. A., 2015, A\&A, 574, A48 

\end{thebibliography}
\end{document}